# Fe I OSCILLATOR STRENGTHS FOR TRANSITIONS FROM HIGH-LYING EVEN-PARITY LEVELS

E. A. Den Hartog[1], M. P. Ruffoni[2], J. E. Lawler[1], J. C. Pickering[2], K. Lind[3] and N. R. Brewer[1]

[1]Department of Physics, University of Wisconsin, Madison, WI 53706, USA
[2]Blackett Laboratory, Imperial College London, London SW7 2BW, UK
[3]Department of Physics and Astronomy, Uppsala University, Box 516, SE-751 20 Uppsala, Sweden
e-mail: eadenhar@wisc.edu

## ABSTRACT

New radiative lifetimes, measured to ±5 % accuracy, are reported for 31 even-parity levels of Fe I ranging from 45061 cm$^{-1}$ to 56842 cm$^{-1}$. These lifetimes have been measured using single-step and two-step time-resolved laser-induced fluorescence on a slow atomic beam of iron atoms. Branching fractions have been attempted for all of these levels, and completed for 20 levels. This set of levels represents an extension of the collaborative work reported in Ruffoni et al. (2014). The radiative lifetimes combined with the branching fractions yields new oscillator strengths for 203 lines of Fe I. Utilizing a 1D-LTE model of the solar photosphere, spectral syntheses for a subset of these lines which are unblended in the solar spectrum yields a mean iron abundance of $<\log[\varepsilon(\text{Fe})]> = 7.45 \pm 0.06$.

## 1. INTRODUCTION

Iron group elements are the heaviest which can be produced through energy-generating nuclear fusion reactions in massive stars and in supernovae, resulting in a 'pile-up' of abundance at what is referred to as the Fe-peak. $^{56}$Fe is particularly abundant because it lies at the maximum of the binding energy per nucleon curve, and thus is the most stable nuclide. In addition to being very abundant, iron also has a complex atomic structure leading to a very rich spectrum throughout the UV, visible and near infrared. It is, therefore, a dominant contributor in the line absorption spectra of most stellar photospheres and is easy to detect in all but a few metal-poor stars. Multiply ionized Fe-peak elements contribute to the opacity and deeper structure of a star. It would be difficult to overstate the importance of iron in astronomical studies. The metallicity of a star is defined as the logarithm of the ratio of iron to hydrogen abundances and iron is used as a reference element in studies of nucleosynthesis and the chemical evolution of the Galaxy. Iron transition probabilities are not only used to determine iron abundances but often other stellar parameters such as the surface gravity and the effective temperature, $T_{eff}$. (see, for example, Gray (2005) Ch. 15)

There is an ongoing need to improve the quality and scope of transition probabilities for the Fe I spectrum. The spectrum of Fe I has been studied in detail by Nave et al. (1994)[1]. This extensive work catalogues nearly 10,000 lines of Fe I from 170 nm to 5 μm, giving energy level classifications for most. In addition she has added 28 new levels and improved the energies of 818 levels. Of the ≈10,000 lines of Fe I available on the NIST Atomic Spectra Database (ASD) website, however, only ≈2400 have measured transition probabilities. These are available from the recent critical compilation by Fuhr & Wiese (2006)[1]. This compilation summarizes in detail the current state of available laboratory transition probabilities for Fe I, and a full discussion will not be presented here. However, the work of two groups is considered particularly noteworthy by the compilers. Those are the very accurate absorption measurements by Blackwell, et al (1979a, 1979b, 1980, 1982a, 1982b, 1986, 1995) for ≈200 transitions and the accurate and comprehensive set of 1814 transition probabilities of O'Brian et al. (1991) most of which were determined by combining measured radiative lifetimes of 186 levels with branching fractions measured using Fourier transform spectroscopy. Using the levels with measured lifetimes, O'Brian et al. also extrapolated level populations in an inductively coupled plasma source to report absolute transition probabilities for an additional 640 transitions from 104 levels.

---
[1] Numerical data from this compilation are available in the NIST database at http://physics.nist.gov/PhysRefData/ASD/index.html

The current study is second in a series on Fe I arising from a collaboration between University of Wisconsin – Madison (UW) and Imperial College, London (IC), with UW contributing the radiative lifetimes measured using time resolved laser induced fluorescence (TR-LIF) and IC contributing the branching fractions from Fourier transform spectra recorded at IC and at the National Institute of Standards and Technology (NIST) We report new radiative lifetimes of 31 high-lying even parity levels, all but three of which are measured for the first time. Branching fractions are attempted for all of these levels, and completed for 20 levels. The lifetimes and branching fractions are combined to determine transition probabilities for 203 lines of the Fe I spectrum. Nine of these levels are ones for which O'Brian et al. (1991) determined transition probabilities from extrapolated level populations. Comparison is made to that technique as well as other data in the Fuhr & Wiese (2006) compilation.

## 2. EXPERIMENTAL PROCEDURES

### 2.1 Radiative Lifetime Measurements

Radiative lifetimes are measured using single-step and two-step laser-induced fluorescence on an atomic beam of iron atoms. These techniques are discussed in detail in the first paper in this series, Ruffoni et al. (2014), so a somewhat abbreviated description is given here. The interested reader is referred to that paper for more detail.

A gas phase sample of iron is produced by sputtering in a hollow cathode discharge operating in ≈50 Pa argon gas, with 30 mA DC current maintaining the discharge between ≈10 A, 10 μs duration pulses. Argon ions accelerated through the cathode fall potential are efficient at sputtering the iron foil which lines the hollow cathode. The iron atoms are differentially pumped through a 1 mm hole in the far end of the cathode, entrained in a flow of argon, into a lower pressure scattering chamber held at ~$10^{-2}$ Pa. Extensive tests have shown this pressure to be low enough to avoid any collisional depopulation of the excited levels under study, which would artificially shorten the fluorescence decay. The iron density is low enough to avoid optical depth problems which would artificially lengthen the fluorescence decay. The resulting 'beam' of iron atoms is weakly collimated and slow moving (≈$5\times10^{+4}$ cm/s).

Four of the levels reported here are measured using single-step laser induced fluorescence. In this technique, a laser beam from a nitrogen laser-pumped dye laser passes through the scattering chamber 1 cm below the bottom of the cathode and is tuned to an appropriate wavelength to selectively excite the level of interest. Narrowband laser excitation eliminates any possibility of cascade radiation from higher lying levels – a problem which was a source of error in older, non-selective excitation techniques – although cascade radiation from lower levels is still a potential problem. The laser used is operated at 30 Hz repetition rate, has pulse duration of ≈3 ns and is tunable over the range 205 nm to 720 nm with the use of a wide variety of dyes and frequency doubling crystals. A delay generator is used to trigger the laser ≈20 μs after the discharge current pulse, which allows for the transit time of the atoms between the bottom of the cathode and the laser interaction region. This point where the beams intersect is at the center of a set of Helmholz coils which cancels the earth's magnetic field. This is necessary to prevent the excited atoms, which have an induced dipole moment along the laser polarization axis, from precessing about the earth's magnetic field resulting in Zeeman quantum beats in the collected fluorescence.

Fluorescence is collected in a direction perpendicular to both the atom and laser beams. The optical train consists of a pair of fused silica lenses which make up an f/1 system. An optical filter, either a broadband colored glass filter or a narrowband multilayer dielectric filter, can be inserted between the lenses where the fluorescence is roughly collimated. Filters are chosen to optimize fluorescence throughput while blocking any cascade radiation from lower levels and reducing or eliminating scattered laser light when possible. The laser – atomic beam interaction region is imaged onto the photocathode of a RCA 1P28A photomultiplier tube (PMT) and the fluorescence signal is recorded with either a boxcar averager for correct line identification or a Tektronix SCD1000 transient digitizer for measuring the exponential decay.

The desired transition for laser excitation is found by recording a LIF spectrum in a 0.5 nm to 1.0 nm range in the vicinity of the line of interest using boxcar averaging. The separations of the lines on this spectrum is then compared to the Fe I, II line list from NIST ASD[1] for identification. Once the line is identified, the laser wavelength is tuned to it, and the time-resolved fluorescence is recorded with the digitizer. The recording of the fluorescence decay begins only after the laser pulse is completely terminated, making deconvolution of the laser excitation and exponential decay unnecessary. An average of 640 fluorescence decays is recorded with the laser tuned to the transition and another 640 traces are averaged with the laser off-line to record the background. The recorded fluorescence is divided into early and late time intervals and the background-subtracted fluorescence is least-square fit to a single exponential in each interval. Comparison of the early and late lifetimes is an easy and sensitive way to ensure that the decay is a clean exponential and all systematic effects are understood and controlled. A set of five lifetimes are averaged for a given set of experimental conditions. The lifetime of each level is measured twice, using different laser excitation when possible, to ensure that the lines used are identified correctly in the experiment and are free from blends.

Two-step laser excitation was required to populate the remaining 27 levels. This technique, in its essentials, is similar to single-step excitation, but using two lasers creates an added level of complexity in the timing of the experiment and more stringent optical filtering of the fluorescence. More details of this technique, particularly the triggering and timing of the experiment, can be found in Ruffoni et al (2014). The two laser beams, each from an independent nitrogen laser-pumped dye laser, enter the scattering chamber at a small angle relative to one another such that they intersect within the viewing volume. The first laser to enter the chamber (laser 1), drives a transition from the ground level or low-lying even parity metastable level to an intermediate odd-parity level. This intermediate level is relatively long-lived (60 ns to 90 ns) in most cases, with one intermediate level as short-lived as ≈10 ns. After the intermediate level has been populated, and before its population can decay away, the second laser (laser 2) enters the viewing volume to excite the level of interest.

Once the wavelength of laser is tuned to populate the intermediate level, it is then left there throughout the measurement. A narrow-band, multilayer dielectric optical filter is put in place which blocks all fluorescence from the intermediate level while allowing fluorescence through from the even-parity level of interest. Laser 2 is then scanned in the vicinity of the 2nd-step transition and a LIF spectrum is recorded. This spectrum typically has one or more single-step transitions as well as the desired two-step transition on it. Once this transition is located it is verified as a two-step transition by blocking each laser independently and making sure that the fluorescence disappears in both cases. Data is then recorded as in the single-step experiment by tuning laser 2 on and off the transition while recording signal and background averages. Unlike the single-step measurement, these lifetimes are measured only once. The use of two narrowband lasers with the fluorescence disappearing when either is blocked, serves the same verification purpose that measuring the lifetime with two different transitions serves in the single-step experiment.

As a final check that the experiment is operating reproducibly and that the systematic effects are well understood and controlled, we routinely measure a set of benchmark lifetimes which are well known from other sources, either from experiments which have smaller and generally different sources of systematic uncertainty than ours or from very accurate theoretical calculations. During the course of the Fe I experiment, we measured the following benchmark lifetimes: $2^2P_{3/2}$ state of $Be^+$ at 8.8519(8) ns (variational method calculation, Yan et al. 1998); the $3^2P_{3/2}$ state of neutral Na at 16.23(1) ns (NIST critical compilation of Kelleher and Podobedova 2008); $4p´[1/2]_1$ state of Ar at 27.85(7) ns (beam-gas-laser-spectroscopy, Volz & Schmoranzer 1998). These benchmarks are measured in the same manner as the Fe I lifetimes except that the cathode lining is different for the $Be^+$ and Na measurements.

The results of our single and two-step lifetime measurements are presented in Table 1 and compared to the limited data available in the literature for these levels. Marek et al. (1979) measured two levels of the $e^5D$ term, 45061 cm$^{-1}$ and 45333 cm$^{-1}$. Their lifetimes are a few percent longer than ours, but still well within the combined uncertainties. O'Brian et al. (1991) measured the lifetime of the $e^7F$ level at 50342 cm$^{-1}$ and we see perfect agreement with that measurement. The O'Brian paper was a collaborative effort between our UW lab, which provided the radiative lifetimes, and colleagues at California Institute of Technology and the National

Solar Observatory, who measured the branching fractions, so agreement with this earlier measurement is reassuring but not surprising.

## 2.2 Branching Fractions and Transition Probabilities

From our recent work for the APOGEE and Gaia-ESO surveys (Ruffoni et al. 2013a, Ruffoni et al. 2014), we have assembled a catalogue of high-resolution, intensity calibrated Fe I line spectra, measured in emission by Fourier transform (FT) spectroscopy in overlapping regions from 1800 cm$^{-1}$ to 35500 cm$^{-1}$. From this catalogue, we have used the two spectra shown in Table 2 to measure branching fractions (BFs) for lines linked to many of the levels listed in Table 1. BF measurements were attempted for all 31 levels, and were successfully completed for 20 of them. With these BFs, we are able to provide experimental Einstein A coefficients for these lines, which are listed in Table 3. These values are needed by astronomers for stellar atmosphere models, and particularly for measurements of stellar metallicity (see Bigot & Thévenin (2006), for example).

Spectrum A was measured between 8200 cm$^{-1}$ and 27500 cm$^{-1}$ on the 2 m FT spectrometer at NIST (Nave et al. 1997), and is described in detail in Ruffoni et al. (2014). The Fe I emission was generated from an iron cathode mounted in a water cooled hollow cathode lamp (HCL) running at a current of 2.0 A in a Ne atmosphere of 370 Pa pressure. The set of scans listed as Spectrum A(1) were measured in a single run using Si photodiodes mounted on both outputs of the spectrometer. The scans listed as Spectrum A(2) were measured in a different run, and were acquired using only one Si photodiode mounted on the unbalanced output of the spectrometer. In both cases, the individual scans listed in the last column of Table 2 were coadded to improve the signal-to-noise ratio of weak lines. The spectrometer response as a function of wavenumber was obtained by measuring the spectrum of a calibrated tungsten (W) halogen lamp with spectral radiance known to ±1.1 % between 250 nm and 2400 nm. W lamp spectra were acquired both before and after measurements of the Fe/Ne HCL spectrum to verify that the spectrometer response remained stable.

Of the 20 upper levels reported in Table 3, 19 contained lines within the range of Spectrum A. For 13 of those levels, Spectrum A(1) alone was used to measure the required BFs. For the remaining 6 levels, some important lines were observed to have a low signal-to-noise ratio in Spectrum A(1). In these cases, Spectrum A(2) was coadded with Spectrum A(1) to improve the signal-to-noise of those lines, and the combined data (A(1) + A(2)) was used to obtain their BFs. For these levels, care was taken to ensure that the relative intensity ratios of the lines were unchanged by having coadded scans acquired during different runs.

Spectrum B was measured between 20000 cm$^{-1}$ and 35500 cm$^{-1}$ on the Imperial College VUV FT spectrometer (Thorne 1996), and is again described in Ruffoni et al. (2014). The Fe I emission was generated from an iron cathode mounted in a HCL running at a current of 700 mA in a Ne atmosphere of 170 Pa pressure to provide reasonable signal-to-noise ratio in the weaker lines while avoiding self-absorption effects in the stronger lines. The spectrometer response function was again obtained from a calibrated W lamp, measured before and after each Fe/Ne HCL measurement. Uncertainties in the relative spectral radiance of the W lamp used at IC, and calibrated by the National Physical Laboratory (NPL), do not exceed ±1.4 % between 410 nm and 800 nm, and rise to ±2.8 % at 300 nm.

The spectrometer response functions used to calibrate the intensities of the lines observed in Spectrum A and B are shown in Figure 1, and were obtained with the aid of the `FAST` package (Ruffoni 2013b). Two separate curves were used to calibrate Spectrum A due to the different detector configurations used when measuring spectrum A(1) and A(2).

Most of the emission lines listed in Table 3 were observed in Spectrum A, which provided a complete set of BFs for almost half of the upper levels studied here. For the remaining levels, significant branches existed at wavenumbers beyond the upper end of Spectrum A, and were instead observed in Spectrum B. In these cases, all the lines belonging to a given upper level were put on a common relative intensity scale by comparing the intensity of lines belonging to the upper level that were observed in both Spectrum A and Spectrum B in the overlap region between 20000 cm$^{-1}$ and 25500 cm$^{-1}$. Spectrum B was then scaled in intensity until the intensity of these 'bridge' lines matched the intensities observed in Spectrum A. If no suitable bridge lines were available for a given upper level, that level was removed from our BF measurements and is thus absent from Table 3. The process of obtaining BFs from overlapping spectra in this way is discussed in detail in Pickering et al. (2001a) and Pickering et al. (2001b).

In all cases, the individual line intensities were obtained by using the `XGremlin` package (Nave et al. 1997) to fit Voigt functions to the observed line profiles. The fit residuals were examined to ensure that each line was free from self-absorption and was not blended with another line. The results were then loaded into the `FAST` package (Ruffoni 2013b), where the relative intensity of each line belonging to a given upper level was measured, providing the BFs shown in Table 3. Given these values, and the lifetimes presented in Table 1, the Einstein coefficient, A, of each line was calculated in the usual way (Huber & Sandeman 1986).

$$A_{ul} = \frac{BF_{ul}}{\tau_u}; \quad \tau_u = \frac{1}{\sum_l A_{ul}} \quad . \tag{1}$$

The subscript $u$ denotes a target upper energy level, and $ul$, a transition from this level to a lower state, $l$. $\tau_u$ is the radiative lifetime of the upper level. The BF$_{ul}$ for a given transition is the ratio of its A$_{ul}$ to the sum of all A$_{ul}$ associated with $u$. This is equivalent to the ratio of observed relative line intensities $I$ in photons/s for these transitions.

$$BF_{ul} = \frac{A_{ul}}{\sum_i A_{ui}} = \frac{I_{ul}}{\sum_i I_{ui}} \quad . \tag{2}$$

This approach does not require any form of equilibrium in the population of different levels, but it is essential that all transitions from $u$ to $l$ be included in the sum over $i$. For any line absent from our spectra for experimental reasons – such as being too weak to be observed, blended, or lying outside of the measured spectral range – a BF value was obtained from previously published laboratory measurements, where available, or derived from the calculations of Kurucz (2007). These were then used to scale the sum over $i$ in Equation 2, and are listed in Table 3 as the "residual" BF for each upper level.

Also shown in Table 3 is the absorption oscillator strength, $f$, of each line, which has been expressed as log($gf$) by convention, where $g$ is the statistical weight of the lower level. These are derived from $A_{ul}$ using the equation

$$\log(gf) = \log[A_{ul} g_u \lambda^2 \times 1.499 \times 10^{-14}] \quad , \tag{3}$$

where $\lambda$ is the wavelength, in nm, of the emission line produced by the electronic transition from level $u$ to level $l$, and $g_u$ is the statistical weight of the upper level.

The sources of uncertainty in measuring and calibrating the intensity of a given line are described in Ruffoni (2013b). The uncertainty in any given BF, $\Delta BF_{ul}$, is given by

$$\left(\frac{\Delta BF_{ul}}{BF_{ul}}\right)^2 = (1 - 2BF_{ul})\left(\frac{\Delta I_{ul}}{I_{ul}}\right)^2 + \sum_{j=1}^{n} BF_{uj}^2 \left(\frac{\Delta I_{uj}}{I_{uj}}\right)^2 \quad , \tag{4}$$

where $I_{ul}$ is the calibrated relative intensity of the emission line associated with the electronic transition from level $u$ to level $l$, and $\Delta I_{ul}$ is the uncertainty in this value. This equation is derived from that given by Sikström et al. (2002) and takes account of the correlation between line intensities, which appear in both the numerator and denominator on the right of Equation 2. From Equation 1, it then follows that the uncertainty in $A_{ul}$, $\Delta A_{ul}$, is

$$\left(\frac{\Delta A_{ul}}{A_{ul}}\right)^2 = \left(\frac{\Delta BF_{ul}}{BF_{ul}}\right)^2 + \left(\frac{\Delta \tau_{ul}}{\tau_{ul}}\right)^2 , \quad (5)$$

where $\Delta \tau_u$ is the 5% uncertainty in the upper level lifetime. This equation can then be used to obtain the uncertainty in log(*gf*).

$$\Delta \log(gf) = \log\left(1 + \frac{\Delta A_{ul}}{A_{ul}}\right). \quad (6)$$

## 3. RESULTS AND COMPARISONS

The results of our BF measurements as well as Einstein A coefficients and oscillator strengths, log(*gf*), are given in Table 3 along with their associated uncertainties for 203 lines of Fe I connected to 20 of the levels listed in Table 1. The table is ordered by increasing upper level energy. The upper level configuration, term and energy as well as lifetime and the percent completeness of the BF analysis appears in a header row followed by the branches associated with that upper level. Also given, where available, is the recommended oscillator strength from the recent critical compilation of Fuhr & Wiese (2006) along with the reference of the original work. In the final column of the table, a letter code is given which is indicative of the method by which the previously published data was put on an absolute scale. An 'L' in this column indicates that the radiative lifetime was measured and used to convert the BF's into transition probabilities. The majority of the lines from three levels, 45333.872 cm$^{-1}$, 45509.149 cm$^{-1}$ and 50342.13 cm$^{-1}$ are scaled in this way. An 'A' in the final column indicates that the previous data arises from an emission measurement from a wall-stabilized arc. This older technique is considerably less accurate than modern BF/lifetime combination data, and these log(*gf*)s were included only for lines for which a more accurate measurement was unavailable. These data are from the work of May et al. (1974) and Bridges & Kornblith (1974) and both were originally normalized to the absolute scale of Bridges & Kornblith. This scale, which was tied to older lifetimes measured using a non-selective excitation technique that was often affected by cascade error, was adjusted for levels higher than 36000 cm$^{-1}$ by Fuhr & Wiese to bring the scale into agreement with the lifetimes of O'Brian et al (1991). The data of Bridges & Kornblith appearing in Table 3 were given a rating of D (±50 %) or D+ by Fuhr & Wiese, and that of May et al. received a rating of D or E (>50 % but within a factor of 3).

The remaining previously published lines in Table 3, indicated by a 'P' in the final column, are from O'Brian et al. (1991). Instead of being put on an absolute scale using radiative lifetimes, however, they extrapolated energy level populations in their inductively coupled plasma source between levels with known radiative lifetimes. Their stated uncertainty in the extrapolated level populations is between 8 % and 9 %, but a comparison with our data suggests this may be an underestimate. Figure 2 shows the difference between our log(*gf*)s and O'Brian's versus upper level energy. Each vertical cluster of points belongs to one upper level. A rectangle encloses each cluster simply to guide the eye in the crowded region above 50000 cm$^{-1}$. The heavy horizontal bar in each cluster represents the calibration offset between our lifetime calibrated log(*gf*)s and the population calibrated data. This calibration offset was obtained for each upper level by taking the logarithm of the sum over the O'Brian Einstein A values for all lines associated with the level multiplied by our lifetime. Perfect agreement between our log(*gf*) values and those of O'Brian would be a difference of zero, indicated by the heavy horizontal line. The heavy dashed lines indicate ±9 % in the *gf*-values, the maximum stated uncertainty in the O'Brian populations, and it is clear that many of the levels lie significantly outside this range.

Figure 3 shows a comparison of our log(*gf*)s with those published in Fuhr & Wiese (2006) versus our log(*gf*)s. For this figure we have rescaled the O'Brian log(*gf*)s by the offsets indicated by the horizontal bars in Figure 2. The wide scatter observed in the May et al (1974) data is not surprising, given the large uncertainties assigned them by Fuhr & Wiese. We see very good agreement with the Bridges & Kornblith (1974) data on four of the five lines that overlap with our study. Figure 3 shows excellent agreement between our study and those of Bard et al. (1991) and Bard et al. (1994), which are scaled using the lifetimes of Marek et al. (1978), with all data agreeing within ±10 %. What is somewhat surprising is the amount of scatter seen in the comparison with the rescaled O'Brian et al. (1991) study. Since it is rescaled with the current set of lifetimes, this comes down to a comparison of the BFs, which are measured using Fourier transform spectroscopy in both studies. Still, the large majority of these lines are in agreement within ±25 %.

## 4. SOLAR SPECTRAL SYNTHESIS

The new atomic data are applied to the determination of the solar iron abundance. The Sun, being well-studied with its high resolution spectrum (Kurucz et al. 1984) and well known fundamental parameters (US Naval Observatory & H.M. Nautical Almanac Office 2013), provides an excellent test-bed in which to assess the new data's impact on stellar syntheses generally. We have determined line-by-line solar Fe abundances for 59 lines selected from Table 3 using both our new log(*gf*)s and the best available from the literature that are not derived from an astrophysical source. This subset of lines, listed in Table 4, was chosen because they were free from blends at the spectral resolution of the Kitt Peak FT Spectrometer (R ≈ 200 000) flux atlas (Kurucz et al. 1984) and have good broadening parameters and continuum placement.

Local thermodynamic equilibrium is assumed for the spectral synthesis, with the one-dimensional, plane-parallel radiative transfer code SME (Valenti & Piskunov 1996), using a MARCS model atmosphere (Gustafsson et al. 2008). We adopted a solar effective temperature $T_{eff}$ = 5777 K, a surface gravity log(grav.) = 4.44, a microturbulence of $\xi_{vmic}$ = 1.0 km s$^{-1}$ and a projected rotational velocity of $v_{rot}\sin(i)$ = 2.0 km s$^{-1}$. The radial-tangential macro-turbulence velocity, $\xi_{vmac}$, was varied between 1.5 and 2.5 km s$^{-1}$ to match the observed profile. The instrumental profile was assumed to be Gaussian. Each line profile was fitted individually using $\chi^2$-minimization between observed and synthetic spectra while varying the iron abundance.

The results of the solar analysis are shown in Figure 4. The line-by-line abundances, log [ε(Fe)],[2] are plotted as a function of log(*gf*) for the previously published data in the top panel and for the new data in the lower panel. The error bars are indicative of the uncertainty contributed from the experimental log(*gf*) values, and do not include other sources of uncertainty in the synthesis process. The new experimental data yield a mean abundance of <log[ε(Fe)]> = 7.45 ± 0.06 dex,[3] in contrast to the data from published sources which yield a mean abundance of <log[ε(Fe)]> = 7.53 ± 0.15 dex. Both the higher average and the greater scatter is driven in large part by the lack of availability of laboratory measurements for some of the lines. The abundance of 7.45 from the new data is in good agreement with recent publications, such as 7.43 ± 0.02 from Bergemann et al. (2012, MARCS, LTE result) and 7.44 ± 0.08 from Ruffoni et al. (2014). This level of agreement is reassuring, and reinforces the overall trustworthiness of the new data set.

## 5. SUMMARY

In conclusion, we report new radiative lifetimes to ±5 % for 31 even-parity levels of Fe I, most of which are measured for the first time. In addition we report new transition probabilities and log(*gf*)s for 203 lines of Fe I associated with 20 of the levels for which lifetimes were measured. Although many of these lines had transition probabilities in publication, our new transition probabilities have significantly lower uncertainties than those in the literature. A solar spectral synthesis of an unblended subset of these lines yields a mean iron abundance of <log[ε(Fe)]> = 7.45 ± 0.06, in excellent agreement with recent values in the literature.

---

[2] We adopt standard stellar spectroscopic notation such that log [ε(Fe)] = $\log_{10}(N_{Fe}/N_H)$ + 12 , where $N_{Fe}$ and $N_H$ are the number densities of iron and hydrogen atoms, respectively.

[3] The unit dex stands for decimal exponent, $x$ dex = $10^x$.

## ACKNOWLEDGEMENTS

We would like to thank Gillian Nave of the Atomic Spectroscopy Group at NIST, Gaithersburg for vital experimental assistance, without which we would have been unable to obtain FT spectra in the infra-red. EDH and JEL acknowledge the support of the US National Science Foundation (NSF) for funding the LIF lifetime measurements under grants AST-0907732 and AST-121105. MPR and JCP would like to thank the UK Science and Technology Facilities Council (STFC) for funding the FT spectroscopy measurements.

## REFERENCES

Anstee S. D., & O'Mara B. J. 1991, MNRAS, 253, 549
Anstee S. D., & O'Mara B. J. 1995, MNRAS, 276, 859
Bard A., Kock A., & Kock M. 1991, A&A, 248, 315
Bard, A., & Kock, M. 1994, A&A, 282, 1014
Bergemann M., Lind K., Collet R., Magic Z., & Asplund M. 2012, MNRAS, 427, 27
Bigot L. & Thévenin F. 2006, MNRAS, 372, 609
Blackwell, D. E., Ibbetson, P. A., Petford, A. D., & Shallis, M. J. 1979, MNRAS, 186, 633
Blackwell, D. E., Petford, A. D., & Shallis, M. J. 1979, MNRAS, 186, 657
Blackwell, D. E., Petford, A. D., Shallis, M. J. & Simmons, G. J. 1980, MNRAS, 191, 445
Blackwell, D. E., Petford, A. D., Shallis, M. J. & Simmons, G. J. 1982, MNRAS, 199, 43
Blackwell, D. E., Petford, A. D., & Simmons, G. J. 1982, MNRAS, 201, 595
Blackwell, D. E., Booth, A. J., Haddock, D. J., Petford, A. D., & Leggett, S. K. 1986, MNRAS, 220, 549
Blackwell, D. E., Smith, G., & Lynas-Gray, A. E. 1995, A&A, 303
Bridges, J. M. & Kornblith, R. L. 1974, ApJ, 192, 793
Fuhr, J. R. & Wiese, W. L. 2006, JPCRD, 35, 1669–1809
Gray, D. F. 2005, The Observation and Analysis of Stellar Photospheres, (3rd ed.; NY, NY: Cambridge Univ. Press)
Gustafsson B., Edvardsson B., Eriksson K., et al. 2008, A&A, 486, 951
Huber, M. C. E., & Sandeman, R. J., 1986, RPPh, 49, 397
Kelleher, D. E., & Podobedova, L. I. 2008, JPCRD, 37, 267
Kurucz, R. L., Furenlid, I., Brault, J., Testerman, L. 1984, National Solar Observatory Atlas: Solar flux atlas from 296 to 1300 nm National Solar Observatory, Sunspot, NM
Kurucz, R. L. 2007, http://kurucz.harvard.edu/atoms/2700/
Marek J., Richter J., & Stahnke H. J. 1979, PhyS, 19, 325
Nave, G., Johansson, S., Learner, R. C. M., Thorne, A. P., & Brault, J. W. 1994, ApJS 94, 221
Nave, G., Sansonetti, C. J., & Griesmann, U. 1997, in OSA Technical Digest Ser. Vol. 3, Fourier Transform Spectroscopy, Optical Society of America, Washington, DC, p. 38
O'Brian, T. R., Wickliffe, M. E., Lawler, J. E., Whaling, W. & Brault, J. W. 1991, JOSAB 8, 1185
Pickering, J. C., Johansson, S., & Smith, P. L. 2001a, A&A, 377, 361
Pickering, J. C., Thorne, A. P., & Perez, R. 2001b, ApJS, 132, 403
Ruffoni, M. P., Allende-Prieto, C., Nave, G., & Pickering, J. C. 2013a, ApJ, 779, 17
Ruffoni, M. P. 2013b, CoPhC, 184, 1770
Ruffoni, M. P., Den Hartog, E. A., Lawler, J. E., et al. 2014, MNRAS, 441, 3127
Sikstrom, C. M., Nilsson, H., Litźen, U., Blom, A. & Lundberg, H. 2002, JQSRT, 74, 355
Thorne, A. P. 1996, PhST, 65, 31
US Naval Observatory, H.M. Nautical Almanac Office, 2013, The Astronomical Almanac for the Year 2014. (Washington, DC: USGPO & London, UK: The Stationery Office)
Valenti J. A., & Piskunov N. 1996, A&AS, 118, 595
Volz U & Schmoranzer H. 1998 in AIP Conf. Proc. 434, *Atomic and Molecular Data and Their Applications*, ed. P. J. Mohr and W. L. Wiese (Woodbury, NY: AIP) 67
Yan, Z-C, Tambasco, M., & Drake, G. W. F. 1998 PhRvA 57, 1652

**Table 1.** Radiative lifetimes for even parity Fe I levels.

| Configuration[a] | Term[a] | J | Upper Level[a] (cm$^{-1}$) | Intermediate Level[a] (cm$^{-1}$) | Laser Wavelengths step 1 (nm) | Laser Wavelengths step 2 (nm) | Observation Wavelength[b] (nm) | Our Lifetime[c] (ns) | Published Lifetime (ns) |
|---|---|---|---|---|---|---|---|---|---|
| | | | | Radiative Lifetimes measured with two-step excitation | | | | | |
| 3d$^6$($^5$D)4s ($^6$D)5s | e$^5$D | 3 | 45061.326 | 26140.177 | 382.444 | 528.362 | 560 | 15.4 | 15.8(0.9)[d] |
| 3d$^6$($^5$D)4s ($^6$D)5s | e$^5$D | 2 | 45333.872 | 26140.177 | 382.444 | 520.859 | 560 | 15.3 | 15.7(0.9)[d] |
| 3d$^6$($^5$D)4s ($^6$D)5s | e$^5$D | 1 | 45509.149 | 26339.694 | 385.637 | 521.518 | 557 | 15.4 | |
| 3d$^7$($^4$F)5s | e$^5$F | 5 | 47005.503 | 25899.987 | 385.991 | 473.677 | 497 | 18.3 | |
| 3d$^7$($^4$F)5s | e$^5$F | 3 | 47755.534 | 26339.696 | 385.637 | 466.813 | 488 | 18.5 | |
| 3d$^7$($^4$F)5s | e$^5$F | 2 | 48036.670 | 26339.696 | 385.637 | 460.765 | 488 | 18.3 | |
| 3d$^6$($^5$D)4s ($^6$D)4d | f$^5$D | 4 | 50423.134 | 25899.987 | 385.991 | 407.663 | 468 | 8.4 | |
| 3d$^6$($^5$D)4s ($^6$D)4d | f$^5$D | 2 | 50698.617 | 26339.694 | 385.637 | 410.411 | 462 | 9.5 | |
| 3d$^6$($^5$D)4s ($^6$D)4d | f$^5$D | 1 | 50880.099 | 26339.694 | 385.637 | 407.376 | 470 | 9.9 | |
| 3d$^6$($^5$D)4s ($^6$D)4d | f$^5$D | 0 | 50981.009 | 26479.379 | 387.857 | 408.021 | 360 | 10.5 | |
| 3d$^6$($^5$D)4s ($^6$D)4d | e$^7$P | 4 | 50475.285 | 25899.987 | 385.991 | 406.798 | 467 | 8.4 | |
| 3d$^6$($^5$D)4s ($^6$D)4d | e$^7$P | 3 | 50611.258 | 25899.987 | 385.991 | 404.559 | 370 | 9.1 | |
| 3d$^6$($^5$D)4s ($^6$D)4d | e$^7$G | 1 | 51566.799 | 26479.379 | 387.857 | 398.493 | 353 | 6.5 | |
| 3d$^6$($^5$D)4s ($^6$D)4d | f$^5$F | 5 | 51103.188 | 26874.548 | 501.207 | 412.618 | 354 | 10.6 | |
| 3d$^6$($^5$D)4s ($^6$D)4d | f$^5$F | 2 | 51705.011 | 26339.694 | 385.637 | 394.127 | 416 | 16.0 | |
| 3d$^6$($^5$D)4s ($^6$D)4d | f$^5$F | 1 | 51754.494 | 26339.694 | 385.637 | 393.360 | 415 | 16.7 | |
| 3d$^6$($^5$D)4s ($^4$D)5s | e$^3$D | 2 | 51739.917 | 26339.694 | 385.637 | 393.586 | 500 | 10.4 | |
| 3d$^6$($^5$D)4s ($^4$D)5s | g$^5$D | 4 | 51350.489 | 25899.987 | 385.991 | 392.808 | 448 | 11.1 | |
| 3d$^6$($^5$D)4s ($^4$D)5s | g$^5$D | 1 | 52214.342 | 26339.694 | 385.637 | 386.369 | 495 | 11.4 | |
| 3d$^6$($^5$D)4s ($^4$D)5s | g$^5$D | 0 | 52257.342 | 26479.379 | 387.857 | 387.818 | 407 | 12.1 | |
| 3d$^6$($^5$D)4s ($^6$D)4d | e$^5$P | 3 | 51837.235 | 25899.987 | 385.991 | 385.437 | 444 | 14.6 | |
| 3d$^6$($^5$D)4s ($^6$D)4d | e$^5$P | 1 | 52019.666 | 26339.694 | 385.637 | 389.298 | 445 | 15.1 | |
| 3d$^7$($^4$F)4d | g$^5$F | 5 | 53061.314 | 26874.548 | 501.207 | 381.764 | 516 | 11.9 | |
| 3d$^7$($^4$F)4d | h$^5$D | 4 | 53155.141 | 26874.548 | 501.207 | 380.401 | 514 | 12.0 | |
| 3d$^7$($^4$F)4d | h$^5$D | 3 | 53545.829 | 26140.177 | 382.444 | 364.784 | 500 | 12.5 | |
| 3d$^7$($^4$F)4d | f$^5$G | 6 | 53169.142 | 26874.548 | 501.207 | 380.198 | 513 | 12.0 | |
| | | | | Radiative Lifetimes measured with single-step excitation | | | | | |
| 3d$^6$($^5$D)4s ($^6$D)4d | e$^7$F | 6 | 50342.126 | | 322.579 | | | 5.6 | 5.6[e] |
| | | | | | 361.016 | | | | |
| 3d$^6$($^5$D)4s ($^6$D)4d | f$^5$D | 3 | 50534.394 | | 322.779 | | | 8.2 | |
| | | | | | 324.820 | | | | |
| 3d$^6$($^5$D)4s ($^6$D)4d | e$^7$G | 7 | 50651.629 | | 357.025 | | | 6.4 | |
| 3d$^6$4s($^6$D)5d | $^7$F | 5 | 56842.729 | | 268.159 | | | 15.9 | |
| | | | | | 301.742 | | | | |

[a] Configurations, terms, level energies and wavelengths are from the Nave et al. (1994).
[b] Fluorescence is observed through ≈10 nm bandpass multi-layer dielectric filters. The filter angle is adjusted to center the bandpass at the indicated wavelength.
[c] ±5% uncertainty.
[d] Marek et al. (1978)
[e] O'Brian et al. (1991) TR-LIF with uncertainty of ±5%.

**Table 2.** FT spectra acquired for BF measurements. These were previously described in Ruffoni et al. (2014).

| Spectrum | Wavenumber Range (cm$^{-1}$) | Detector | Filter | Resolution (cm$^{-1}$) | Spectrum Filename[a] |
|---|---|---|---|---|---|
| A(1) (NIST) | 8200 - 25500 | 2× Si photodiode | None | 0.02 | Fe080311.001 to .003 (110 coadds) |
| A(2) (NIST) | 8200 - 25500 | 1× Si photodiode | None | 0.02 | Fe080411_B.001 to .003 (110 coadds) |
| B (IC) | 20000 - 35500 | Hamamatsu R11568 PMT | Schott BG3 | 0.037 | Fe130610.002 to .047 (96 coadds) |

[a] The named spectra were coadded to improve the signal-to-noise ratio of the weak lines.

**Table 3.** Experimental branching fractions, transition probabilities and log(*gf*)s for 20 even-parity levels of Fe I.

| Lower Level (cm$^{-1}$) | J | $\lambda_{air}$ (Å) | BF | Δ BF (%) | $A_{ul}$ (10$^6$ s$^{-1}$) | This Experiment log(*gf*) | Published log(*gf*) | Ref.[a] |
|---|---|---|---|---|---|---|---|---|
| | | Upper Level: 45061.326 cm$^{-1}$; 3d$^6$($^5$D)4s ($^6$D)5s e$^5$D$_3$; τ = 15.4 ns ± 5 %; 97 % complete | | | | | | |
| 40594.429 | 4 | 22380.797 | | | | | -0.46 ± 0.05 | OB91 P |
| 40231.333 | 2 | 20698.313 | | | | | -1.05 ± 0.06 | OB91 P |
| 39969.850 | 3 | 19635.309 | | | | | -0.68 ± 0.05 | OB91 P |
| 39625.801 | 4 | 18392.461 | | | | | -1.09 ± 0.07 | OB91 P |
| 37157.564 | 2 | 12648.742 | | | | | -1.14 ± 0.05 | OB91 P |
| 36766.964 | 3 | 12053.083 | | | | | -1.54 ± 0.12 | OB91 P |
| 34039.514 | 4 | 9070.4342 | 0.003 | 6.5 | 0.202 | -1.76 ± 0.03 | | |
| 29469.022 | 2 | 6411.6493 | 0.093 | 1.6 | 6.034 | -0.59 ± 0.02 | -0.72 ± 0.04 | OB91 P |
| 29056.322 | 3 | 6246.3188 | 0.065 | 1.3 | 4.195 | -0.77 ± 0.02 | -0.88 ± 0.04 | OB91 P |
| 27559.581 | 2 | 5712.1316 | 0.004 | 15.6 | 0.286 | -2.01 ± 0.07 | -1.99 ± 0.03 | Ba91 L |
| 27394.689 | 3 | 5658.8164 | 0.079 | 1.8 | 5.129 | -0.76 ± 0.02 | -0.84 ± 0.04 | OB91 P |
| 27166.818 | 4 | 5586.7559 | 0.369 | 0.7 | 23.964 | -0.11 ± 0.02 | -0.14 ± 0.04 | OB91 P |
| 26339.694 | 2 | 5339.9294 | 0.120 | 0.9 | 7.810 | -0.63 ± 0.02 | -0.72 ± 0.05 | OB91 P |
| 26140.177 | 3 | 5283.6210 | 0.187 | 0.9 | 12.153 | -0.45 ± 0.02 | -0.53 ± 0.04 | OB91 P |
| 25899.987 | 4 | 5217.3893 | 0.046 | 1.0 | 2.979 | -1.07 ± 0.02 | -1.16 ± 0.04 | OB91 P |
| 24180.860 | 3 | 4787.8269 | 0.002 | 14.0 | 0.101 | -2.62 ± 0.06 | -2.60 ± 0.12 | OB91 P |
| 23711.454 | 4 | 4682.5605 | | | | | -3.07 ± 0.30 | Ma74 A |
| 23192.498 | 2 | 4571.4378 | | | | | -3.21 ± 0.30 | Ma74 A |
| 23110.937 | 3 | 4554.4512 | 0.001 | 21.6 | 0.071 | -2.81 ± 0.09 | -2.99 ± 0.18 | Ma74 A |
| 19912.494 | 2 | 3975.2055 | | | | | -2.70 ± 0.18 | Ma74 A |
| | | Upper Level: 45333.872 cm$^{-1}$; 3d$^6$($^5$D)4s ($^6$D)5s e$^5$D$_2$; τ = 15.3 ns ± 5 %; 97 % complete | | | | | | |
| 29732.734 | 1 | 6408.0184 | 0.051 | 2.7 | 3.361 | -0.99 ± 0.02 | -1.02 ± 0.03 | Ba91 L |
| 29469.022 | 2 | 6301.5012 | 0.100 | 2.2 | 6.545 | -0.71 ± 0.02 | -0.72 ± 0.03 | Ba91 L |
| 29056.322 | 3 | 6141.7320 | 0.019 | 4.4 | 1.234 | -1.46 ± 0.03 | -1.46 ± 0.03 | Ba91 L |
| 27666.346 | 1 | 5658.5317 | 0.009 | 4.6 | 0.586 | -1.85 ± 0.03 | -1.86 ± 0.03 | Ba91 L |
| 27559.581 | 2 | 5624.5422 | 0.112 | 1.3 | 7.331 | -0.76 ± 0.02 | -0.76 ± 0.03 | Ba91 L |
| 27394.689 | 3 | 5572.8424 | 0.346 | 0.9 | 22.631 | -0.28 ± 0.02 | -0.28 ± 0.03 | Ba91 L |
| 26479.379 | 1 | 5302.3030 | 0.136 | 1.3 | 8.891 | -0.73 ± 0.02 | -0.72 ± 0.03 | Ba91 L |
| 26339.694 | 2 | 5263.3063 | 0.100 | 1.6 | 6.560 | -0.87 ± 0.02 | -0.88 ± 0.03 | Ba91 L |
| 26140.177 | 3 | 5208.5940 | 0.098 | 1.3 | 6.414 | -0.89 ± 0.02 | -0.90 ± 0.03 | Ba91 L |
| 24180.860 | 3 | 4726.1370 | 0.001 | 23.4 | 0.076 | -2.89 ± 0.09 | -3.19 ± 0.30 | Ma74 A |
| 23244.836 | 1 | 4525.8635 | | | | | -3.14 ± 0.30 | Ma74 A |

| | | | | | | | | | |
|---|---|---|---|---|---|---|---|---|---|
| 23192.498 | 2 | 4515.1667 | | | | | -3.17 ± 0.30 | | Ma74 A |

Upper Level: 45509.149 cm⁻¹;  3d⁶(⁵D)4s (⁶D)5s e⁵D₁ ;  τ = 15.4 ns ± 5 %;  98 % complete

| | | | | | | | |
|---|---|---|---|---|---|---|---|
| 34547.209 | 2 | 9119.9705 | 0.001 | 31.0 | 0.064 | -2.62 ± 0.12 | |
| 29732.734 | 1 | 6336.8243 | 0.122 | 2.9 | 7.906 | -0.85 ± 0.02 | -0.86 ± 0.03 | Ba94 L |
| 29469.022 | 2 | 6232.6412 | 0.051 | 3.6 | 3.288 | -1.24 ± 0.03 | -1.22 ± 0.03 | Ba94 L |
| 27666.346 | 1 | 5602.9451 | 0.142 | 1.5 | 9.226 | -0.89 ± 0.02 | -0.85 ± 0.03 | Ba94 L |
| 27559.581 | 2 | 5569.6181 | 0.337 | 1.1 | 21.886 | -0.52 ± 0.02 | -0.49 ± 0.03 | Ba94 L |
| 26550.477 | 0 | 5273.1636 | 0.120 | 1.7 | 7.799 | -1.01 ± 0.02 | -0.99 ± 0.03 | Ba94 L |
| 26479.379 | 1 | 5253.4617 | 0.033 | 2.0 | 2.114 | -1.58 ± 0.02 | -1.57 ± 0.03 | Ba94 L |
| 26339.694 | 2 | 5215.1806 | 0.172 | 1.8 | 11.185 | -0.86 ± 0.02 | -0.87 ± 0.03 | Ba94 L |
| 23270.382 | 0 | 4495.3936 | 0.001 | 39.1 | 0.045 | -3.39 ± 0.14 | |

Upper Level: 47005.503 cm⁻¹;  3d⁷(⁴F)5s e⁵F₅ ;  τ = 18.3 ns ± 5 %;  98 % complete

| | | | | | | | |
|---|---|---|---|---|---|---|---|
| 40257.311 | 5 | 14814.735 | | | | | -1.03 ± 0.08 | OB91 P |
| 35379.206 | 5 | 8598.8302 | 0.009 | 3.7 | 0.515 | -1.20 ± 0.03 | -1.09 ± 0.04 | OB91 P |
| 34843.955 | 6 | 8220.3790 | 0.327 | 0.7 | 17.893 | 0.30 ± 0.02 | 0.28 ± 0.05 | OB91 P |
| 34039.514 | 4 | 7710.3645 | 0.016 | 4.7 | 0.865 | -1.07 ± 0.03 | -1.11 ± 0.06 | OB91 P |
| 33695.395 | 5 | 7511.0205 | 0.257 | 0.8 | 14.022 | 0.12 ± 0.02 | 0.10 ± 0.04 | OB91 P |
| 33095.939 | 4 | 7187.3180 | 0.174 | 1.0 | 9.502 | -0.09 ± 0.02 | -0.15 ± 0.05 | OB91 P |
| 27166.818 | 4 | 5039.2520 | 0.013 | 3.6 | 0.718 | -1.52 ± 0.03 | -1.57 ± 0.04 | OB91 P |
| 26874.548 | 5 | 4966.0889 | 0.073 | 1.4 | 3.980 | -0.79 ± 0.02 | -0.87 ± 0.06 | OB91 P |
| 25899.987 | 4 | 4736.7734 | 0.106 | 1.4 | 5.781 | -0.67 ± 0.02 | -0.75 ± 0.04 | OB91 P |

Upper Level: 47755.534 cm⁻¹;  3d⁷(⁴F)5s e⁵F₃ ;  τ = 18.5 ns ± 5 %;  97 % complete

| | | | | | | | |
|---|---|---|---|---|---|---|---|
| 38678.036 | 2 | 11013.237 | 0.007 | 11.5 | 0.369 | -1.33 ± 0.05 | |
| 38175.352 | 3 | 10435.356 | 0.001 | 22.5 | 0.075 | -2.07 ± 0.09 | |
| 37521.158 | 2 | 9768.3155 | 0.002 | 38.8 | 0.080 | -2.09 ± 0.14 | |
| 37162.744 | 3 | 9437.7961 | 0.007 | 7.2 | 0.369 | -1.46 ± 0.04 | |
| 35767.562 | 4 | 8339.4039 | 0.073 | 3.7 | 3.931 | -0.54 ± 0.03 | |
| 35611.623 | 3 | 8232.3178 | 0.037 | 2.9 | 1.971 | -0.85 ± 0.02 | |
| 35257.322 | 4 | 7998.9458 | 0.263 | 1.7 | 14.193 | -0.02 ± 0.02 | |
| 34547.209 | 2 | 7568.8999 | 0.041 | 3.3 | 2.205 | -0.88 ± 0.03 | |
| 34328.750 | 3 | 7445.7508 | 0.186 | 2.0 | 10.043 | -0.23 ± 0.02 | |
| 34039.514 | 4 | 7288.7385 | 0.019 | 5.6 | 1.043 | -1.24 ± 0.03 | |
| 33946.931 | 2 | 7239.8676 | 0.015 | 28.2 | 0.829 | -1.34 ± 0.11 | |
| 33801.570 | 2 | 7164.4486 | 0.099 | 2.4 | 5.345 | -0.54 ± 0.02 | |
| 33507.121 | 3 | 7016.3920 | 0.030 | 4.0 | 1.632 | -1.07 ± 0.03 | -1.21 ± 0.18 | Ma74 A |
| 27559.581 | 2 | 4950.1060 | 0.023 | 3.0 | 1.227 | -1.50 ± 0.03 | |
| 27394.689 | 3 | 4910.0169 | 0.039 | 2.6 | 2.082 | -1.28 ± 0.02 | |
| 27166.818 | 4 | 4855.6732 | 0.018 | 3.5 | 0.946 | -1.63 ± 0.03 | |
| 26339.694 | 2 | 4668.1344 | 0.068 | 3.1 | 3.654 | -1.08 ± 0.03 | |
| 26140.177 | 3 | 4625.0453 | 0.044 | 2.7 | 2.385 | -1.27 ± 0.02 | |
| 25899.987 | 4 | 4574.2162 | 0.004 | 14.5 | 0.203 | -2.35 ± 0.06 | -2.45 ± 0.30 | Ma74 A |

Upper Level: 48036.670 cm⁻¹;  3d⁷(⁴F)5s e⁵F₂ ;  τ = 18.3 ns ± 5 %;  97 % complete

| | | | | | | |
|---|---|---|---|---|---|---|
| 37521.158 | 2 | 9507.1551 | 0.003 | 8.7 | 0.167 | -1.95 ± 0.04 |
| 36079.370 | 3 | 8360.7956 | 0.028 | 6.8 | 1.524 | -1.10 ± 0.04 |
| 35856.400 | 2 | 8207.7429 | 0.040 | 3.9 | 2.170 | -0.96 ± 0.03 |
| 35611.623 | 3 | 8046.0479 | 0.300 | 1.9 | 16.401 | -0.10 ± 0.02 |
| 34692.146 | 1 | 7491.6486 | 0.038 | 5.0 | 2.088 | -1.06 ± 0.03 |
| 34547.209 | 2 | 7411.1544 | 0.165 | 2.4 | 9.039 | -0.43 ± 0.02 |

| | | | | | | | | |
|---|---|---|---|---|---|---|---|---|
| 34328.750 | 3 | 7293.0454 | 0.051 | 3.6 | 2.761 | -0.96 ± 0.03 | | |
| 34017.101 | 1 | 7130.9221 | 0.099 | 2.9 | 5.387 | -0.69 ± 0.02 | -0.80 ± 0.18 | Ma74 A |
| 33801.570 | 2 | 7022.9539 | 0.044 | 5.3 | 2.422 | -1.05 ± 0.03 | -1.20 ± 0.18 | Ma74 A |
| 27666.346 | 1 | 4907.7318 | 0.020 | 5.8 | 1.101 | -1.70 ± 0.03 | | |
| 27559.581 | 2 | 4882.1434 | 0.034 | 4.2 | 1.839 | -1.48 ± 0.03 | | |
| 27394.689 | 3 | 4843.1438 | 0.023 | 4.5 | 1.262 | -1.65 ± 0.03 | -1.79 ± 0.30 | Ma74 A |
| 26479.379 | 1 | 4637.5034 | 0.059 | 3.1 | 3.205 | -1.29 ± 0.03 | -1.34 ± 0.18 | Ma74 A |
| 26339.694 | 2 | 4607.6469 | 0.054 | 3.0 | 2.950 | -1.33 ± 0.03 | | |
| 26140.177 | 3 | 4565.6619 | 0.008 | 9.1 | 0.434 | -2.17 ± 0.04 | -2.20 ± 0.30 | Ma74 A |

Upper Level: 50342.126 cm$^{-1}$; 3d$^6$($^5$D)4s ($^6$D)4d e$^7$F$_6$ ; $\tau$ = 5.6 ns ± 5 %; 100 % complete

| | | | | | | | | |
|---|---|---|---|---|---|---|---|---|
| 40257.311 | 5 | 9913.1793 | 0.001 | 12.0 | 0.094 | -1.74 ± 0.05 | | |
| 26874.548 | 5 | 4259.9992 | 0.006 | 10.3 | 1.119 | -1.40 ± 0.05 | -1.26 ± 0.18 | OB91 L |
| 22650.414 | 6 | 3610.1591 | 0.290 | 11.2 | 51.844 | 0.12 ± 0.05 | 0.18 ± 0.05 | OB91 L |
| 19350.890 | 5 | 3225.7872 | 0.702 | 4.7 | 125.404 | 0.41 ± 0.03 | 0.38 ± 0.03 | OB91 L |

Upper Level: 50423.134 cm$^{-1}$; 3d$^6$($^5$D)4s ($^6$D)4d f$^5$D$_4$ ; $\tau$ = 8.4 ns ± 5 %; 96 % complete

| | | | | | | | | |
|---|---|---|---|---|---|---|---|---|
| 43499.502 | 4 | 14439.341 | | | | | -1.14 ± 0.13 | OB91 P |
| 39625.801 | 4 | 9259.0055 | 0.014 | 3.4 | 1.707 | -0.71 ± 0.03 | -0.75 ± 0.10 | OB91 P |
| 36766.964 | 3 | 7320.6826 | | | | | -1.16 ± 0.08 | OB91 P |
| 29056.322 | 3 | 4678.8458 | 0.060 | 2.8 | 7.094 | -0.68 ± 0.02 | -0.83 ± 0.04 | OB91 P |
| 26874.548 | 5 | 4245.3444 | 0.013 | 7.7 | 1.538 | -1.43 ± 0.04 | -1.61 ± 0.06 | OB91 P |
| 25899.987 | 4 | 4076.6291 | 0.097 | 2.7 | 11.536 | -0.59 ± 0.02 | -0.53 ± 0.04 | OB91 P |
| 24180.860 | 3 | 3809.5643 | 0.014 | 20.7 | 1.626 | -1.50 ± 0.08 | -1.55 ± 0.06 | OB91 P |
| 23711.454 | 4 | 3742.6166 | 0.069 | 6.5 | 8.208 | -0.81 ± 0.03 | -0.89 ± 0.04 | OB91 P |
| 22996.672 | 4 | 3645.0748 | 0.023 | 9.3 | 2.706 | -1.31 ± 0.04 | -1.28 ± 0.12 | OB91 P |
| 22845.867 | 5 | 3625.1417 | 0.079 | 3.8 | 9.389 | -0.78 ± 0.03 | -0.84 ± 0.05 | OB91 P |
| 19757.031 | 3 | 3259.9895 | 0.031 | 8.6 | 3.718 | -1.27 ± 0.04 | -1.37 ± 0.04 | OB91 P |
| 19562.438 | 4 | 3239.4329 | 0.369 | 3.9 | 43.955 | -0.21 ± 0.03 | -0.38 ± 0.04 | OB91 P |
| 19350.890 | 5 | 3217.3772 | 0.189 | 5.3 | 22.448 | -0.50 ± 0.03 | -0.68 ± 0.04 | OB91 P |

Upper Level: 50475.285 cm$^{-1}$; 3d$^6$($^5$D)4s ($^6$D)4d e$^7$P$_4$ ; $\tau$ = 8.4 ns ± 5 %; 97 % complete

| | | | | | | | | |
|---|---|---|---|---|---|---|---|---|
| 43499.502 | 4 | 14331.390 | | | | | -0.94 ± 0.13 | OB91 P |
| 40421.935 | 4 | 9944.2057 | 0.003 | 4.3 | 0.396 | -1.28 ± 0.03 | | |
| 40257.311 | 5 | 9783.9910 | 0.004 | 22.4 | 0.457 | -1.23 ± 0.09 | | |
| 39625.801 | 4 | 9214.4977 | 0.013 | 3.1 | 1.518 | -0.76 ± 0.03 | | |
| 36766.964 | 3 | 7292.8294 | 0.008 | 14.0 | 0.955 | -1.16 ± 0.06 | -1.10 ± 0.07 | OB91 P |
| 29056.322 | 3 | 4667.4531 | 0.055 | 3.0 | 6.573 | -0.71 ± 0.03 | -0.75 ± 0.04 | OB91 P |
| 27166.818 | 4 | 4289.0824 | 0.023 | 3.9 | 2.745 | -1.17 ± 0.03 | | |
| 26140.177 | 3 | 4108.1330 | 0.004 | 34.7 | 0.442 | -2.00 ± 0.13 | -2.10 ± 0.30 | Ma74 |
| 25899.987 | 4 | 4067.9777 | 0.111 | 3.0 | 13.150 | -0.53 ± 0.03 | -0.47 ± 0.04 | OB91 P |
| 23711.454 | 4 | 3735.3238 | 0.212 | 6.5 | 25.285 | -0.32 ± 0.03 | -0.29 ± 0.04 | OB91 P |
| 22845.867 | 5 | 3618.2990 | 0.048 | 9.5 | 5.708 | -1.00 ± 0.04 | -1.06 ± 0.06 | OB91 P |
| 19757.031 | 3 | 3254.4569 | 0.002 | 27.1 | 0.289 | -2.38 ± 0.11 | | |
| 19562.438 | 4 | 3233.9675 | 0.188 | 5.3 | 22.332 | -0.50 ± 0.03 | -0.53 ± 0.04 | OB91 P |
| 19350.890 | 5 | 3211.9873 | 0.298 | 4.5 | 35.503 | -0.31 ± 0.03 | -0.19 ± 0.04 | OB91 P |

Upper Level: 50534.394 cm$^{-1}$; 3d$^6$($^5$D)4s ($^6$D)4d f$^5$D$_3$ ; $\tau$ = 8.2 ns ± 5 %; 86 % complete

| | | | | | | | | |
|---|---|---|---|---|---|---|---|---|
| 39969.850 | 3 | 9463.0287 | 0.006 | 7.0 | 0.717 | -1.17 ± 0.04 | | |
| 39625.801 | 4 | 9164.5715 | 0.006 | 7.6 | 0.748 | -1.18 ± 0.04 | | |
| 36766.964 | 3 | 7261.5140 | 0.006 | 18.9 | 0.778 | -1.37 ± 0.08 | | |
| 29469.022 | 2 | 4745.8001 | 0.020 | 6.9 | 2.404 | -1.25 ± 0.04 | -1.27 ± 0.06 | OB91 P |

| | | | | | | | | | |
|---|---|---|---|---|---|---|---|---|---|
| 29056.322 | 3 | 4654.6050 | 0.026 | 4.2 | 3.216 | -1.14 ± 0.03 | -1.08 ± 0.04 | OB91 | P |
| 27166.818 | 4 | 4278.2314 | 0.011 | 10.0 | 1.380 | -1.58 ± 0.05 | -1.70 ± 0.18 | Ma74 | A |
| 26140.177 | 3 | 4098.1758 | 0.050 | 5.5 | 6.093 | -0.97 ± 0.03 | -0.88 ± 0.04 | OB91 | P |
| 25899.987 | 4 | 4058.2170 | 0.031 | 6.4 | 3.802 | -1.18 ± 0.03 | -1.11 ± 0.05 | OB91 | P |
| 24180.860 | 3 | 3793.4806 | 0.040 | 7.2 | 4.909 | -1.13 ± 0.04 | -0.92 ± 0.04 | OB91 | P |
| 23711.454 | 4 | 3727.0924 | | | | | -0.60 ± 0.06 | OB91 | P |
| 22996.672 | 4 | 3630.3478 | 0.089 | 12.5 | 10.879 | -0.82 ± 0.06 | -0.84 ± 0.04 | OB91 | P |
| 19757.031 | 3 | 3248.2042 | 0.159 | 6.6 | 19.360 | -0.67 ± 0.03 | -0.67 ± 0.04 | OB91 | P |
| 19562.438 | 4 | 3227.7955 | 0.415 | 4.3 | 50.598 | -0.26 ± 0.03 | -0.27 ± 0.04 | OB91 | P |

Upper Level: 50651.629 cm$^{-1}$;  3d$^6$($^5$D)4s ($^6$D)4d e$^7$G$_7$ ;  $\tau$ = 6.4 ns ± 5 %;  100 % complete

| | | | | | | | | | |
|---|---|---|---|---|---|---|---|---|---|
| 22650.414 | 6 | 3570.2542 | 1.000 | 0.0 | 156.208 | 0.65 ± 0.02 | | | |

Upper Level: 50698.617 cm$^{-1}$;  3d$^6$($^5$D)4s ($^6$D)4d f$^5$D$_2$ ;  $\tau$ = 9.5 ns ± 5 %;  94 % complete

| | | | | | | | | | |
|---|---|---|---|---|---|---|---|---|---|
| 40842.151 | 3 | 10142.843 | 0.003 | 11.8 | 0.304 | -1.63 ± 0.05 | | | |
| 40231.333 | 2 | 9550.9568 | 0.004 | 11.2 | 0.426 | -1.54 ± 0.05 | | | |
| 39969.850 | 3 | 9318.1787 | 0.012 | 8.7 | 1.209 | -1.10 ± 0.04 | | | |
| 36766.964 | 3 | 7175.9251 | 0.013 | 16.6 | 1.400 | -1.27 ± 0.07 | | | |
| 33507.121 | 3 | 5815.2178 | | | | | -2.58 ± 0.30 | Ma74 | A |
| 29732.734 | 1 | 4768.3203 | 0.022 | 7.4 | 2.283 | -1.41 ± 0.04 | | | |
| 29469.022 | 2 | 4709.0881 | 0.027 | 7.6 | 2.889 | -1.32 ± 0.04 | | | |
| 29056.322 | 3 | 4619.2880 | 0.052 | 6.2 | 5.505 | -1.06 ± 0.03 | -1.08 ± 0.18 | Ma74 | A |
| 27394.689 | 3 | 4289.9146 | 0.013 | 16.2 | 1.327 | -1.74 ± 0.07 | | | |
| 26479.379 | 1 | 4127.7843 | 0.018 | 8.0 | 1.933 | -1.61 ± 0.04 | | | |
| 26339.694 | 2 | 4104.1136 | 0.040 | 7.0 | 4.224 | -1.27 ± 0.04 | | | |
| 26140.177 | 3 | 4070.7707 | 0.104 | 6.0 | 10.935 | -0.87 ± 0.03 | -0.85 ± 0.18 | Br74 | A |
| 24180.860 | 3 | 3769.9874 | 0.076 | 8.0 | 8.013 | -1.07 ± 0.04 | | | |
| 23110.937 | 3 | 3623.7731 | 0.081 | 9.4 | 8.477 | -1.08 ± 0.04 | | | |
| 19912.494 | 2 | 3247.2801 | 0.086 | 9.8 | 8.997 | -1.15 ± 0.05 | | | |
| 19757.031 | 3 | 3230.9636 | 0.393 | 5.0 | 41.410 | -0.49 ± 0.03 | -0.54 ± 0.18 | Br74 | A |

Upper Level: 50880.099 cm$^{-1}$;  3d$^6$($^5$D)4s ($^6$D)4d f$^5$D$_1$ ;  $\tau$ = 9.9 ns ± 5 %;  86 % complete

| | | | | | | | | | |
|---|---|---|---|---|---|---|---|---|---|
| 40404.515 | 1 | 9543.3912 | 0.010 | 24.8 | 1.030 | -1.38 ± 0.10 | | | |
| 40231.333 | 2 | 9388.1837 | 0.015 | 6.7 | 1.511 | -1.22 ± 0.04 | | | |
| 29732.734 | 1 | 4727.3946 | 0.087 | 6.2 | 8.753 | -1.06 ± 0.03 | -1.16 ± 0.05 | OB91 | P |
| 29469.022 | 2 | 4669.1711 | 0.057 | 6.6 | 5.780 | -1.25 ± 0.03 | -1.21 ± 0.05 | OB91 | P |
| 27559.581 | 2 | 4286.8645 | 0.008 | 21.9 | 0.765 | -2.20 ± 0.09 | | | |
| 26550.477 | 0 | 4109.0561 | 0.037 | 11.5 | 3.732 | -1.55 ± 0.05 | -1.56 ± 0.07 | OB91 | P |
| 26479.379 | 1 | 4097.0834 | 0.023 | 14.9 | 2.270 | -1.77 ± 0.06 | -1.65 ± 0.30 | Ma74 | A |
| 26339.694 | 2 | 4073.7623 | 0.142 | 4.7 | 14.324 | -0.97 ± 0.03 | -0.90 ± 0.04 | OB91 | P |
| 24506.915 | 2 | 3790.6547 | 0.046 | 15.2 | 4.591 | -1.53 ± 0.06 | -1.55 ± 0.08 | OB91 | P |
| 23192.498 | 2 | 3610.6946 | 0.083 | 12.3 | 8.428 | -1.31 ± 0.05 | -1.21 ± 0.05 | OB91 | P |
| 20019.634 | 1 | 3239.4574 | | | | | -1.50 ± 0.05 | OB91 | P |
| 19912.494 | 2 | 3228.2490 | 0.351 | 6.9 | 35.465 | -0.78 ± 0.04 | -0.76 ± 0.05 | OB91 | P |

Upper Level: 50981.009 cm$^{-1}$;  3d$^6$($^5$D)4s ($^6$D)4d f$^5$D$_0$ ;  $\tau$ = 10.5 ns ± 5 %;  96 % complete

| | | | | | | | | | |
|---|---|---|---|---|---|---|---|---|---|
| 40404.515 | 1 | 9452.3359 | 0.023 | 12.5 | 2.171 | -1.54 ± 0.06 | | | |
| 29732.734 | 1 | 4704.9481 | 0.152 | 12.3 | 14.484 | -1.32 ± 0.05 | -1.53 ± 0.18 | Ma74 | A |
| 27666.346 | 1 | 4287.9385 | 0.043 | 27.2 | 4.070 | -1.95 ± 0.11 | | | |
| 26479.379 | 1 | 4080.2092 | 0.278 | 16.9 | 26.510 | -1.18 ± 0.07 | -1.23 ± 0.18 | Br74 | A |
| 23244.836 | 1 | 3604.3716 | 0.138 | 22.8 | 13.179 | -1.59 ± 0.09 | | | |
| 20019.634 | 1 | 3228.8987 | 0.329 | 18.2 | 31.362 | -1.31 ± 0.08 | | | |
| | | Upper Level: 51350.489 cm$^{-1}$; 3d$^6$($^5$D)4s ($^4$D)5s g$^5$D$_4$ ; τ = 11.1 ns ± 5 %; 99 % complete | | | | | | | |
| 40594.429 | 4 | 9294.5351 | 0.004 | 16.5 | 0.377 | -1.36 ± 0.07 | | | |
| 40257.311 | 5 | 9012.0745 | 0.090 | 2.5 | 8.090 | -0.05 ± 0.02 | -0.31 ± 0.09 | OB91 | P |
| 39969.850 | 3 | 8784.4405 | 0.008 | 31.5 | 0.704 | -1.14 ± 0.12 | | | |
| 39625.801 | 4 | 8526.6690 | 0.029 | 5.5 | 2.639 | -0.59 ± 0.03 | -0.76 ± 0.09 | OB91 | P |
| 36766.964 | 3 | 6855.1621 | 0.056 | 2.7 | 5.054 | -0.49 ± 0.02 | -0.74 ± 0.09 | OB91 | P |
| 34039.514 | 4 | 5775.0806 | 0.021 | 4.8 | 1.850 | -1.08 ± 0.03 | -1.30 ± 0.06 | OB91 | P |
| 33695.395 | 5 | 5662.5162 | 0.100 | 2.4 | 8.984 | -0.41 ± 0.02 | -0.57 ± 0.04 | OB91 | P |
| 33507.121 | 3 | 5602.7673 | 0.030 | 3.3 | 2.680 | -0.95 ± 0.03 | -1.14 ± 0.04 | OB91 | P |
| 33095.939 | 4 | 5476.5642 | 0.145 | 2.2 | 13.096 | -0.28 ± 0.02 | -0.45 ± 0.04 | OB91 | P |
| 29056.322 | 3 | 4484.2198 | 0.094 | 2.5 | 8.439 | -0.64 ± 0.02 | -0.86 ± 0.04 | OB91 | P |
| 27166.818 | 4 | 4133.8557 | 0.036 | 5.2 | 3.203 | -1.13 ± 0.03 | -1.31 ± 0.05 | OB91 | P |
| 26874.548 | 5 | 4084.4915 | 0.144 | 2.3 | 12.936 | -0.54 ± 0.02 | -0.71 ± 0.06 | OB91 | P |
| 26140.177 | 3 | 3965.5088 | 0.015 | 6.8 | 1.316 | -1.55 ± 0.04 | -1.78 ± 0.07 | OB91 | P |
| 25899.987 | 4 | 3928.0829 | 0.091 | 5.6 | 8.163 | -0.77 ± 0.03 | -0.93 ± 0.04 | OB91 | P |
| 23110.937 | 3 | 3540.1211 | 0.128 | 2.9 | 11.532 | -0.71 ± 0.02 | -0.80 ± 0.05 | OB91 | P |
| | | Upper Level: 51566.799 cm$^{-1}$; 3d$^6$($^5$D)4s ($^6$D)4d e$^7$G$_1$ ; τ = 6.5 ns ± 5 %; 99 % complete | | | | | | | |
| 26550.477 | 0 | 3996.2623 | 0.004 | 17.4 | 0.624 | -2.35 ± 0.07 | | | |
| 26479.379 | 1 | 3984.9340 | 0.006 | 26.1 | 0.985 | -2.15 ± 0.10 | | | |
| 23270.382 | 0 | 3533.0066 | 0.494 | 1.4 | 76.065 | -0.37 ± 0.02 | -0.32 ± 0.04 | OB91 | P |
| 23244.836 | 1 | 3529.8198 | 0.436 | 1.5 | 67.019 | -0.43 ± 0.02 | -0.36 ± 0.04 | OB91 | P |
| 23192.498 | 2 | 3523.3087 | 0.051 | 14.9 | 7.870 | -1.36 ± 0.06 | -1.23 ± 0.05 | OB91 | P |
| | | Upper Level: 51705.011 cm$^{-1}$; 3d$^6$($^5$D)4s ($^6$D)4d f$^5$F$_2$ ; τ = 16.0 ns ± 5 %; 95 % complete | | | | | | | |
| 41130.596 | 1 | 9454.1951 | 0.059 | 6.2 | 3.686 | -0.61 ± 0.03 | | | |
| 41018.048 | 2 | 9354.6304 | 0.016 | 8.4 | 1.014 | -1.18 ± 0.04 | | | |
| 40404.515 | 1 | 8846.7407 | 0.075 | 7.6 | 4.659 | -0.56 ± 0.04 | | | |
| 40231.333 | 2 | 8713.2034 | 0.050 | 7.9 | 3.116 | -0.75 ± 0.04 | | | |
| 29469.022 | 2 | 4495.9531 | 0.021 | 14.8 | 1.330 | -1.70 ± 0.06 | -1.69 ± 0.18 | Ma74 | A |
| 27666.346 | 1 | 4158.7924 | 0.247 | 4.9 | 15.455 | -0.70 ± 0.03 | -0.70 ± 0.18 | Ma74 | A |
| 27559.581 | 2 | 4140.4024 | 0.031 | 17.0 | 1.940 | -1.60 ± 0.07 | | | |
| 27394.689 | 3 | 4112.3185 | 0.035 | 16.6 | 2.160 | -1.56 ± 0.07 | -1.72 ± 0.18 | Ma74 | A |
| 26479.379 | 1 | 3963.1005 | 0.283 | 4.9 | 17.685 | -0.68 ± 0.03 | -0.76 ± 0.18 | Br74 | A |
| 26339.694 | 2 | 3941.2753 | 0.128 | 7.7 | 8.000 | -1.03 ± 0.04 | -0.98 ± 0.18 | Ma74 | A |
| | | Upper Level: 51739.917 cm$^{-1}$; 3d$^6$($^5$D)4s ($^4$D)5s e$^3$D$_2$ ; τ = 10.4 ns ± 5 %; 96 % complete | | | | | | | |
| 41018.048 | 2 | 9324.1745 | 0.007 | 9.9 | 0.629 | -1.39 ± 0.05 | | | |
| 40842.151 | 3 | 9173.6758 | 0.013 | 4.1 | 1.236 | -1.11 ± 0.03 | | | |
| 38678.036 | 2 | 7653.7596 | 0.040 | 6.3 | 3.852 | -0.77 ± 0.03 | -0.89 ± 0.07 | OB91 | P |
| 37162.744 | 3 | 6858.1498 | 0.038 | 4.6 | 3.609 | -0.90 ± 0.03 | -0.93 ± 0.05 | OB91 | P |
| 34547.209 | 2 | 5814.8075 | | | | | -1.94 ± 0.30 | Ma74 | A |
| 34362.871 | 1 | 5753.1227 | 0.102 | 1.9 | 9.759 | -0.62 ± 0.02 | -0.69 ± 0.04 | OB91 | P |
| 34328.750 | 3 | 5741.8484 | | | | | -1.67 ± 0.08 | OB91 | P |
| 33946.931 | 2 | 5618.6327 | 0.025 | 5.5 | 2.406 | -1.25 ± 0.03 | -1.28 ± 0.05 | OB91 | P |
| 33801.570 | 2 | 5573.1024 | 0.023 | 6.8 | 2.189 | -1.29 ± 0.04 | -1.32 ± 0.05 | OB91 | P |
| 33507.121 | 3 | 5483.0988 | 0.019 | 4.8 | 1.807 | -1.39 ± 0.03 | -1.41 ± 0.05 | OB91 | P |
| 32133.989 | 2 | 5099.0773 | 0.022 | 4.7 | 2.101 | -1.39 ± 0.03 | -1.27 ± 0.04 | OB91 | P |
| 31937.323 | 1 | 5048.4361 | 0.055 | 2.3 | 5.250 | -1.00 ± 0.02 | -1.03 ± 0.04 | OB91 | P |
| 31805.069 | 3 | 5014.9425 | 0.364 | 1.1 | 34.982 | -0.18 ± 0.02 | -0.30 ± 0.05 | OB91 | P |
| 31686.349 | 2 | 4985.2529 | 0.202 | 1.4 | 19.428 | -0.44 ± 0.02 | -0.56 ± 0.04 | OB91 | P |
| 31322.611 | 3 | 4896.4385 | 0.008 | 11.0 | 0.721 | -1.89 ± 0.05 | -2.02 ± 0.30 | Ma74 | A |
| 29469.022 | 2 | 4488.9069 | 0.013 | 9.0 | 1.260 | -1.72 ± 0.04 | -1.83 ± 0.08 | OB91 | P |
| 27559.581 | 2 | 4134.4207 | | | | | | | |
| 27394.689 | 3 | 4106.4229 | 0.026 | 8.3 | 2.530 | -1.50 ± 0.04 | -1.50 ± 0.06 | OB91 | P |

Upper Level: 51754.494 cm$^{-1}$; 3d$^6$($^5$D)4s ($^6$D)4d f$^5$F$_1$; $\tau$ = 16.7 ns ± 5 %; 96 % complete

| | | | | | | | | | |
|---|---|---|---|---|---|---|---|---|---|
| 41130.596 | 1 | 9410.1583 | 0.053 | 8.6 | 3.175 | -0.90 ± 0.04 | | | |
| 41018.048 | 2 | 9311.5149 | 0.019 | 13.4 | 1.120 | -1.36 ± 0.06 | | | |
| 40491.281 | 0 | 8876.0241 | 0.042 | 12.3 | 2.510 | -1.05 ± 0.05 | | | |
| 40404.515 | 1 | 8808.1710 | 0.058 | 14.4 | 3.480 | -0.92 ± 0.06 | | | |
| 29469.022 | 2 | 4485.9725 | | | | | -2.32 ± 0.30 | Ma74 | A |
| 27666.346 | 1 | 4150.2491 | 0.138 | 8.7 | 8.279 | -1.19 ± 0.04 | -1.23 ± 0.18 | Ma74 | A |
| 27559.581 | 2 | 4131.9354 | 0.050 | 24.1 | 2.967 | -1.64 ± 0.10 | | | |
| 26550.477 | 0 | 3966.4995 | 0.254 | 7.5 | 15.221 | -0.97 ± 0.04 | | | |
| 26479.379 | 1 | 3955.3413 | 0.212 | 8.8 | 12.688 | -1.05 ± 0.04 | -0.98 ± 0.18 | Ma74 | A |
| 23270.382 | 0 | 3509.7258 | 0.074 | 31.4 | 4.413 | -1.61 ± 0.12 | | | |
| 23244.836 | 1 | 3506.5803 | 0.063 | 30.1 | 3.781 | -1.68 ± 0.12 | | | |

Upper Level: 53169.142 cm$^{-1}$; 3d$^7$($^4$F)4d f$^5$G$_6$; $\tau$ = 12.0 ns ± 5 %; 93 % complete

| | | | | | | | | | |
|---|---|---|---|---|---|---|---|---|---|
| 42784.349 | 6 | 9626.8254 | 0.008 | 5.1 | 0.673 | -0.92 ± 0.03 | | | |
| 35379.206 | 5 | 5619.5954 | | | | | -1.67 ± 0.30 | Ma74 | A |
| 34843.955 | 6 | 5455.4544 | 0.382 | 1.1 | 31.791 | 0.27 ± 0.02 | | | |
| 33695.395 | 5 | 5133.6885 | 0.540 | 0.8 | 45.021 | 0.36 ± 0.02 | | | |
| 26874.548 | 5 | 3801.9840 | | | | | -0.98 ± 0.30 | Ma74 | A |

Note – Table 3 is also available in machine-readable format organized by increasing wavelength.
[a] The acronyms in the reference column have the following meaning: OB91 – O'Brian et al. (1991); Ma74 – May et al. (1974); Ba91 – Bard et al. (1991); Ba94 – Bard & Kock (1994); Br74 – Bridges & Kornblith (1974). The letter following the reference acronym indicates the method by which the published data was put on an absolute scale as described in the text.

**Table 4.** Lines from Table 3 for which solar line profiles have been synthesized.

| $\lambda_{air}$ (Å) | This experiment | | | Previously published | | | VdW[c] Parameter | log[$\varepsilon$(Fe)] | |
|---|---|---|---|---|---|---|---|---|---|
| | log(gf) | ± | unc. | log(gf) | ± | unc.[a] Ref.[b] | | This experiment | Previous work |
| 4073.7623 | -0.97 | ± | 0.02 | -0.90 | ± | 0.04 OB91 | 789.294 | 7.49 | 7.42 |
| 4080.2092 | -1.18 | ± | 0.07 | -1.23 | ± | 0.18 Br74 | 798.292 | 7.38 | 7.43 |
| 4150.2491 | -1.26 | ± | 0.03 | -1.23 | ± | 0.18 Ma74 | 882.275 | 7.44 | 7.41 |
| 4158.7924 | -0.70 | ± | 0.02 | -0.70 | ± | 0.18 Ma74 | 874.275 | 7.28 | 7.28 |
| 4484.2198 | -0.64 | ± | 0.02 | -0.86 | ± | 0.04 OB91 | 749.253 | 7.35 | 7.57 |
| 4495.3936 | -3.39 | ± | 0.14 | -1.69 | | Ma74 | 869.275 | 7.45 | 5.75 |
| 4574.2162 | -2.35 | ± | 0.06 | -2.45 | ± | 0.30 Ma74 | 886.226 | 7.46 | 7.56 |
| 4619.2880 | -1.06 | ± | 0.03 | -1.08 | ± | 0.18 Ma74 | 764.292 | 7.47 | 7.50 |
| 4625.0453 | -1.27 | ± | 0.02 | -1.34 | | Ku07 | 886.226 | 7.46 | 7.53 |
| 4637.5034 | -1.29 | ± | 0.02 | -1.34 | ± | 0.18 Ma74 | 917.226 | 7.51 | 7.56 |
| 4669.1711 | -1.25 | ± | 0.03 | -1.21 | ± | 0.05 OB91 | 778.287 | 7.52 | 7.48 |
| 4678.8458 | -0.68 | ± | 0.02 | -0.83 | ± | 0.04 OB91 | 734.298 | 7.50 | 7.65 |
| 4704.9481 | -1.32 | ± | 0.05 | -1.53 | ± | 0.18 Ma74 | 786.284 | 7.42 | 7.63 |
| 4726.1370 | -2.89 | ± | 0.09 | -3.19 | ± | 0.30 Ma74 | 836.226 | 7.25 | 7.55 |
| 4736.7734 | -0.67 | ± | 0.01 | -0.75 | ± | 0.04 OB91 | 820.231 | 7.52 | 7.60 |

| λ (Å) | log(gf) | ± | σ | log(gf) ref | ± | σ | Ref[b] | VdW[c] | log ε☉ | log ε |
|---|---|---|---|---|---|---|---|---|---|---|
| 4787.8269 | -2.62 | ± | 0.06 | -2.60 | ± | 0.12 | OB91 | 818.227 | 7.48 | 7.47 |
| 4882.1434 | -1.48 | ± | 0.02 | -2.56 | | | Ku07 | 910.227 | 7.43 | 8.50 |
| 4907.7318 | -1.70 | ± | 0.03 | -1.60 | | | Ku07 | 909.227 | 7.40 | 7.30 |
| 4966.0889 | -0.79 | ± | 0.01 | -0.87 | ± | 0.06 | OB91 | 821.235 | 7.48 | 7.56 |
| 4985.2529 | -0.44 | ± | 0.01 | -0.56 | ± | 0.04 | OB91 | 742.240 | 7.35 | 7.47 |
| 5014.9425 | -0.18 | ± | 0.01 | -0.30 | ± | 0.05 | OB91 | 734.239 | 7.42 | 7.54 |
| 5048.4361 | -1.00 | ± | 0.02 | -1.03 | ± | 0.04 | OB91 | 732.238 | 7.40 | 7.43 |
| 5133.6885 | 0.36 | ± | 0.01 | 0.14 | | | Ku07 | 811.278 | 7.32 | 7.54 |
| 5217.3893 | -1.07 | ± | 0.01 | -1.16 | ± | 0.04 | OB91 | 815.232 | 7.44 | 7.53 |
| 5253.4617 | -1.58 | ± | 0.01 | -1.57 | ± | 0.03 | Ba94 | 849.229 | 7.45 | 7.44 |
| 5273.1636 | -1.01 | ± | 0.01 | -0.99 | ± | 0.03 | Ba94 | 849.230 | 7.43 | 7.41 |
| 5283.6210 | -0.45 | ± | 0.01 | -0.53 | ± | 0.04 | OB91 | 815.233 | 7.47 | 7.55 |
| 5302.3030 | -0.73 | ± | 0.01 | -0.72 | ± | 0.03 | Ba91 | 835.231 | 7.50 | 7.49 |
| 5339.9294 | -0.63 | ± | 0.01 | -0.72 | ± | 0.05 | OB91 | 815.234 | 7.48 | 7.57 |
| 5483.0988 | -1.39 | ± | 0.02 | -1.41 | ± | 0.05 | OB91 | 737.241 | 7.44 | 7.46 |
| 5569.6181 | -0.52 | ± | 0.01 | -0.49 | ± | 0.03 | Ba94 | 848.233 | 7.47 | 7.44 |
| 5572.8424 | -0.28 | ± | 0.01 | -0.28 | ± | 0.03 | Ba91 | 835.235 | 7.45 | 7.45 |
| 5618.6327 | -1.25 | ± | 0.03 | -1.28 | ± | 0.05 | OB91 | 732.214 | 7.46 | 7.49 |
| 5624.5422 | -0.76 | ± | 0.01 | -0.76 | ± | 0.03 | Ba91 | 835.235 | 7.51 | 7.51 |
| 5662.5162 | -0.41 | ± | 0.02 | -0.57 | ± | 0.04 | OB91 | 724.235 | 7.44 | 7.60 |
| 5712.1316 | -2.01 | ± | 0.06 | -1.99 | ± | 0.03 | Ba91 | 817.240 | 7.42 | 7.40 |
| 5753.1227 | -0.62 | ± | 0.01 | -0.69 | ± | 0.04 | OB91 | 741.211 | 7.40 | 7.47 |
| 5775.0806 | -1.08 | ± | 0.02 | -1.30 | ± | 0.06 | OB91 | 720.231 | 7.46 | 7.68 |
| 6246.3188 | -0.77 | ± | 0.01 | -0.88 | ± | 0.04 | OB91 | 820.246 | 7.47 | 7.58 |
| 6301.5012 | -0.71 | ± | 0.01 | -0.72 | ± | 0.03 | Ba91 | 832.243 | 7.47 | 7.48 |
| 6336.8243 | -0.85 | ± | 0.02 | -0.86 | ± | 0.03 | Ba94 | 845.240 | 7.47 | 7.48 |
| 6411.6493 | -0.59 | ± | 0.01 | -0.72 | ± | 0.04 | OB91 | 820.247 | 7.47 | 7.60 |
| 6858.1498 | -0.90 | ± | 0.02 | -0.93 | ± | 0.05 | OB91 | 765.211 | 7.44 | 7.48 |
| 7022.9539 | -1.05 | ± | 0.03 | -1.20 | ± | 0.18 | Ma74 | 912.245 | 7.43 | 7.58 |
| 7130.9221 | -0.69 | ± | 0.02 | -0.80 | ± | 0.18 | Ma74 | 914.246 | 7.45 | 7.56 |
| 7288.7385 | -1.24 | ± | 0.03 | -1.03 | | | Ku07 | -7.55 | 7.38 | 7.17 |
| 7491.6486 | -1.06 | ± | 0.02 | -0.90 | | | Ku07 | -7.55 | 7.57 | 7.41 |
| 7568.8999 | -0.88 | ± | 0.02 | -0.77 | | | Ku07 | -7.55 | 7.54 | 7.44 |
| 7710.3645 | -1.07 | ± | 0.02 | -1.11 | ± | 0.06 | OB91 | -7.55 | 7.48 | 7.52 |
| 8207.7429 | -0.96 | ± | 0.02 | -0.85 | | | Ku07 | -7.55 | 7.53 | 7.42 |
| 8360.7956 | -1.10 | ± | 0.03 | -1.67 | | | Ku07 | -7.55 | 7.49 | 8.07 |
| 8526.6690 | -0.59 | ± | 0.03 | -0.76 | ± | 0.09 | OB91 | -7.53 | 7.50 | 7.67 |
| 8784.4405 | -1.14 | ± | 0.12 | -1.61 | | | Ku07 | -7.53 | 7.47 | 7.95 |
| 8846.7407 | -0.56 | ± | 0.03 | -0.78 | | | Ku07 | -7.50 | 7.41 | 7.62 |
| 8876.0241 | -1.10 | ± | 0.04 | -1.06 | | | Ku07 | -7.49 | 7.55 | 7.51 |
| 9070.4342 | -1.76 | ± | 0.03 | -2.08 | | | Ku07 | -7.54 | 7.30 | 7.62 |
| 9119.9705 | -2.62 | ± | 0.12 | -2.54 | | | Ku07 | -7.54 | 7.38 | 7.29 |
| 9783.9910 | -1.23 | ± | 0.09 | -1.74 | | | Ku07 | -7.54 | 7.48 | 8.00 |
| 9944.2057 | -1.28 | ± | 0.02 | -1.29 | | | Ku07 | -7.54 | 7.54 | 7.55 |

[a] Uncertainties are only available for experimentally measured log(*gf*) values.
[b] Reference acronyms are the same as for Table 3. In addition Ku07 stands for Kurucz et al. (2007)
[c] Van der Waals broadening parameter. Values greater than zero were obtained from Anstee, Barklem & O'Mara (ABO) theory (Anstee & O'Mara 1991, 1995) and are expressed in the standard packed notation where the integer component is the broadening cross section, $\sigma$, in atomic units, and the decimal component is the dimensionless velocity parameter, $\alpha$. Values less than zero are the log of the VdW broadening parameter, $\gamma_6$ (rad s$^{-1}$), per unit perturber number density, N (cm$^{-3}$), at 10000 K (i.e. log[$\gamma_6$/N] in units of rad s$^{-1}$ cm$^3$). These were used only when ABO data were unavailable. See Gray (2005) for more details.

**Figure 1.** Instrument response functions used to intensity calibrate the two FT spectra listed in Table 2.

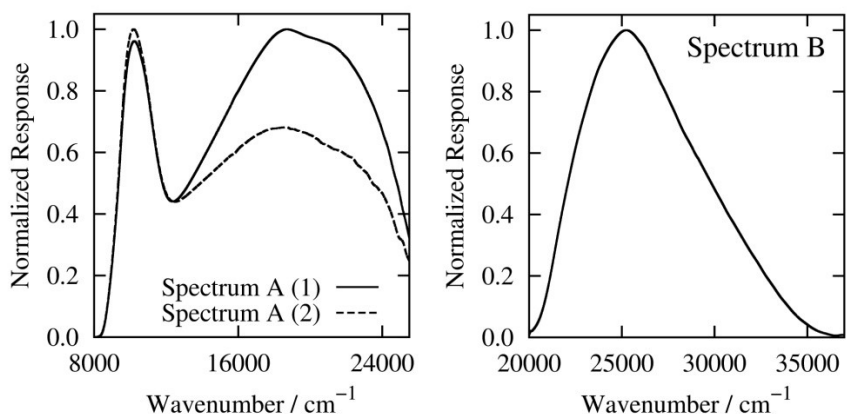

**Figure 2**. Comparison of new log(*gf*)s to those of O'Brian et al. (1991) which were scaled using upper level population extrapolations. Perfect agreement is a difference of zero, and is marked by the heavy solid horizontal line. The heavy dashed horizontal lines either side of zero indicate ± 9% difference in the *gf*-values, the stated uncertainty in the O'Brian level populations. The data are plotted versus upper level energy so that all the points belonging to one level form a vertical cluster. The rectangles superposed over each cluster are simply an aid to guide the eye. The heavy horizontal bar within each rectangle represents the calibration offset for each upper level as discussed in the text.

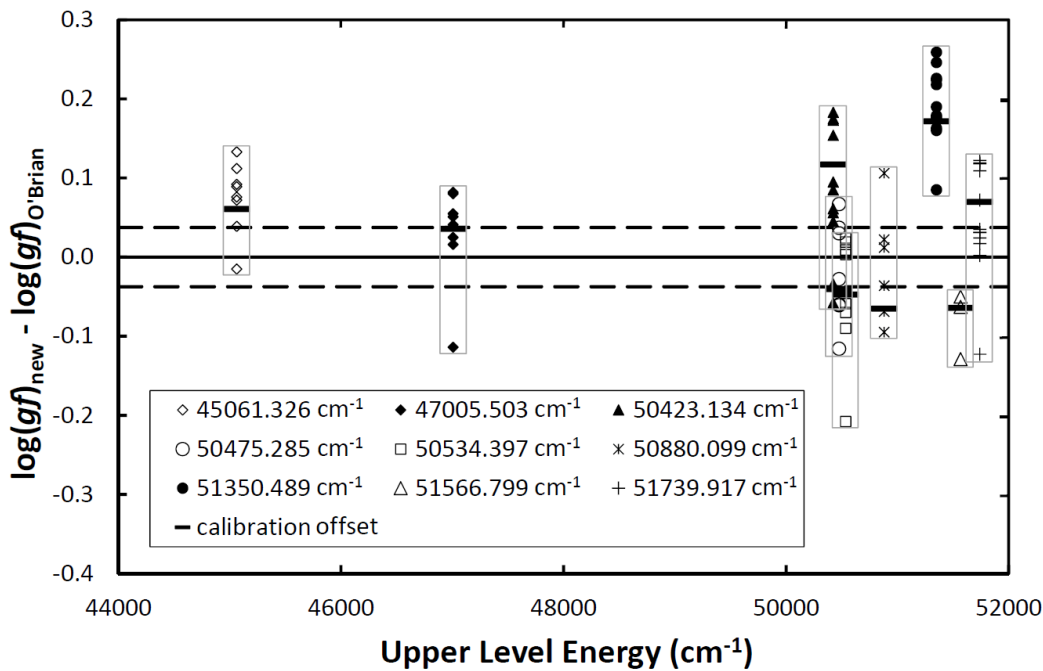

**Figure 3**. Comparison between the log(*gf*)s from this study and those contained in the compilation of Fuhr & Wiese (2006). The May et al. (1974) and Bridges & Kornblith (1974) were rescaled by Fuhr & Wiese (2006) using lifetimes from O'Brian et al. (1991). The O'Brian et al. (1991) results from level population extrapolations are rescaled here to our lifetimes (see discussion in text). The central solid line represents perfect agreement, and the dashed lines represent the uncertainty in the log(*gf*)s as defined in the text for the indicated percent uncertainties in the Einstein A coefficients.

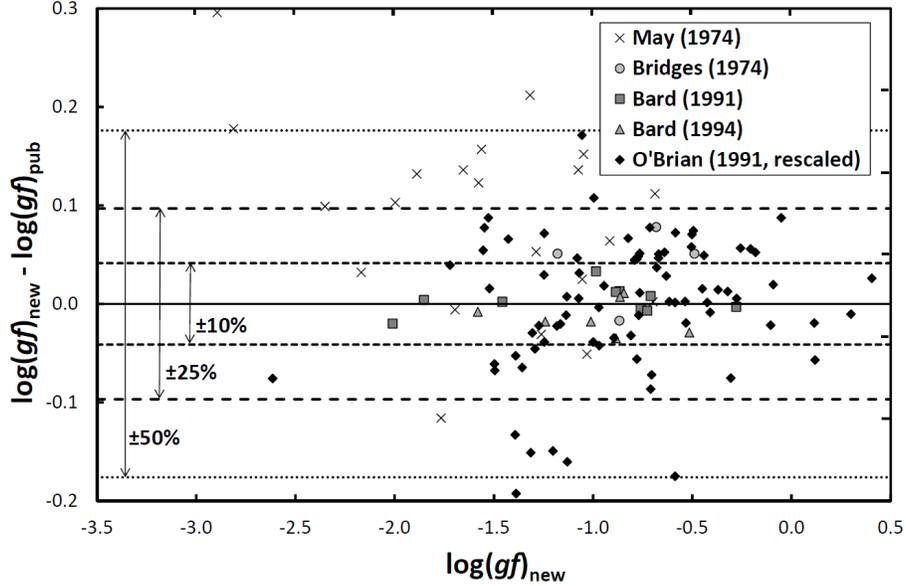

**Fig. 4** Solar iron abundance, log[ε(Fe)], obtained from the synthesis of individual lines listed in Table 4 using the log(*gf*) values from this work, log(*gf*)$_{new}$, and the best previously published values, log(*gf*)$_{pub}$. Only those lines that appear unblended in the solar spectrum are included. The two points with attached arrows in the upper pane are discrepant semi-empirical values that lie outside the plotted range at the values shown to their left. These were not included in the calculation of the average log[ε(Fe)]$_{pub}$. The solid horizontal lines in each figure indicate the average abundance and the dashed horizontal lines indicate the scatter. The error bars indicate only the uncertainty from the experimental log(*gf*)s and do not capture the uncertainties associated with the solar atmospheric modeling and spectral synthesis.

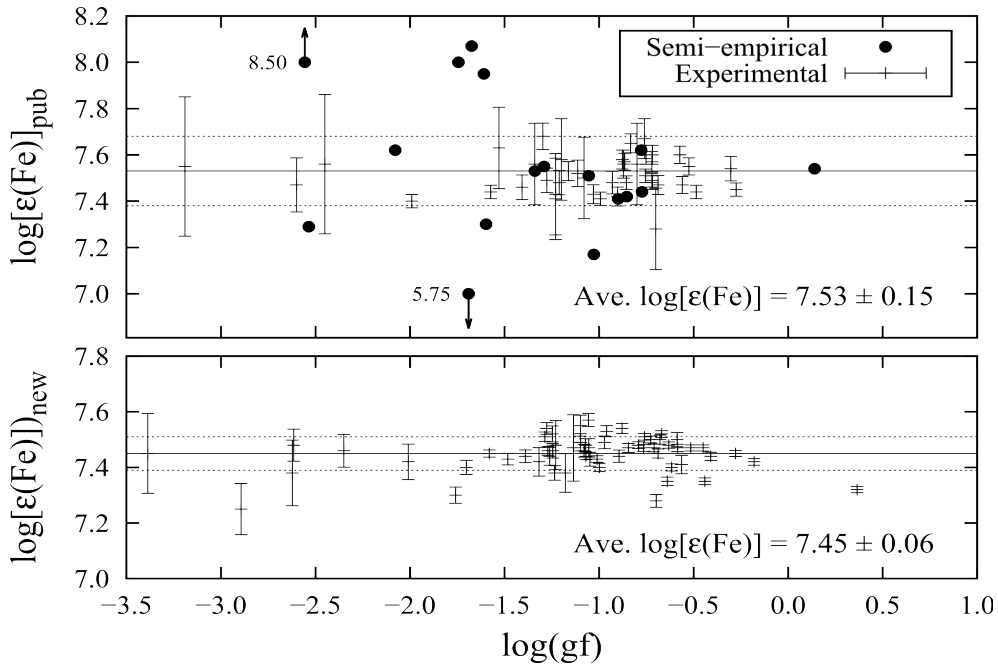